\documentclass[
aps,%
12pt,%
final,%
notitlepage,%
oneside,%
onecolumn,%
nobibnotes,%
nofootinbib,
superscriptaddress,%
noshowpacs,%
centertags]%
{revtex4}
\RequirePackage{graphicx}

\begin{document}
\title{Chiral thermodynamics of dense hadronic matter}

\author{\firstname{Chihiro} \surname{Sasaki}}
\email{sasaki@fias.uni-frankfurt.de}
\affiliation{Frankfurt Institute for Advanced Studies,
D-60438 Frankfurt am Main,
Germany}

\begin{abstract}
We discuss phases of hot and dense hadronic matter using chiral Lagrangians.
A two-flavored parity doublet model constrained by the nuclear matter ground 
state predicts chiral symmetry restoration. The model thermodynamics is
shown within the mean field approximation.
A field-theoretical constraint on possible phases from the anomaly matching
is also discussed.
\end{abstract}

\maketitle

\section{Parity doubled nucleons}
\label{sec:int}

Model studies of hot and dense matter have suggested a rich phase structure 
of QCD at temperatures and quark chemical potentials of order 
$\Lambda_{\rm QCD}$. Our knowledge on the phase structure however remains 
limited and the description of strongly interacting matter does not 
reach a consensus yet~\cite{qmproc}. 
In particular, properties of baryons near the chiral symmetry restoration are
poorly understood.
The realistic modeling of dense baryonic matter must take into
account the existence of the nuclear matter saturation point, i.e. the bound
state at baryon density $\rho_0 = 0.16$ fm$^{-3}$, like in Walecka type
models~\cite{walecka}.
Several chiral models with pure hadronic degrees of freedom~\cite{other,nnjl}
have been constructed in such a way that the nuclear matter has the true 
ground state. An alternative approach is to describe a nucleon as
a dynamical bound-state of a diquark and a quark~\cite{bentz}.

In the mirror assignment of chirality to nucleons~\cite{dk,mirror},
dynamical chiral symmetry breaking generates a mass difference between parity 
partners and the chiral symmetry restoration does not necessarily dictate the 
chiral partners to be massless. Mirror baryons embedded in linear and non-linear 
chiral Lagrangians have been applied to study their phenomenology 
in vacuum~\cite{dk,mirror,lsma1}, nuclear matter~\cite{hp,pdm} 
and neutron starts~\cite{astro}.
Identifying the true parity partner of 
a nucleon is also an issue. In the mirror models $N(1535)$ is usually taken to 
be the negative parity state. This choice however fails to reproduce the decay 
width to a nucleon and $\eta$. This might indicate another negative parity 
state lighter than the $N(1535)$~\cite{pdm}, which has not been observed so far.

\begin{figure}[h]
\begin{center}
\includegraphics[width=8cm]{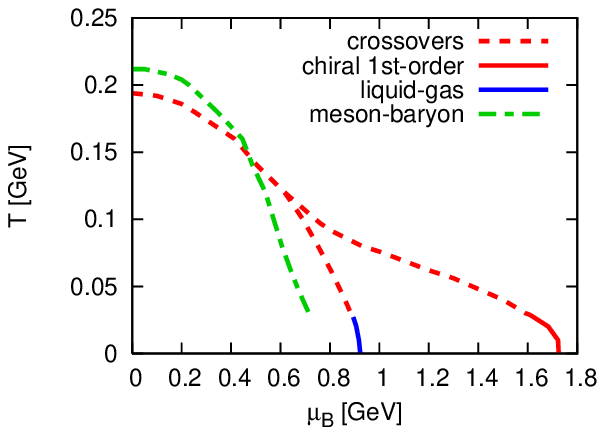}
\includegraphics[width=8cm]{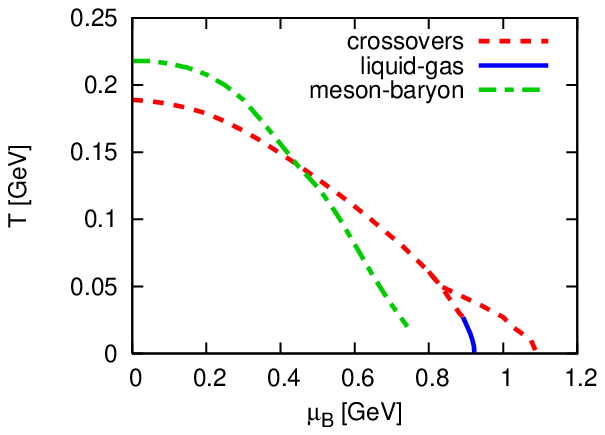}
\caption{
The phase diagram in the parity doublet model~\cite{pdmour}.
The mass of the negative parity nucleon was taken to be $1.5$ GeV (left)
and $1.2$ GeV (right).
}
\label{pdm}
\end{center}
\end{figure}

The parity doublet model has been applied to a hot and dense hadronic 
matter and the phase structure of a chiral symmetry restoration as well as 
a liquid-gas transition of nuclear matter was explored~\cite{pdmour}. 
In Fig.~\ref{pdm} we show the phase diagram for two different masses of
the negative parity state, $m_{N-}=1.5$ GeV and $1.2$ GeV. The latter
is considered to be an phenomenological option.
At zero temperature the system experiences a first-order liquid-gas
transition at $\mu_B = 923$ MeV and the baryon density shows a jump from
zero to a finite value $\rho \neq 0$. 
Roughly speaking chiral symmetry restoration occurs when the baryon chemical 
potential reaches the mass of the negative parity state, $\mu_B \sim m_{N-}$.
The order of chiral phase transition and its location depend on the set of 
parameters, especially on mass of the negative parity state. 
If we take the most frequently used value
$m_{N-}=1500$ MeV, then in addition to the nuclear liquid-gas phase transition
we obtain a weak first-order chiral transition 
at $\rho \sim 10\,\rho_0$. 
With a lower mass $m_{N-}=1200$ MeV 
we get no true chiral phase transition but only a crossover at much lower density
$\rho \sim 3\,\rho_0$.
The liquid-gas transition survives up to $T = 27$ MeV. Above this temperature
there is no sharp phase transition but the order parameter is still attracted 
by the critical point:
the order parameter typically shows a double-step structure and
this makes an additional crossover line terminating at the liquid-gas 
critical point. Another crossover line corresponding to the chiral symmetry 
restoration follows 
the steepest descent of the second reduction in $\langle \sigma \rangle$. 
With increasing
temperature the two crossover lines become closer and finally merge.

In contrast, the trajectory of a meson-to-baryon ``transition'' defined 
from the ratio of particle number densities is basically driven by
the density effect with the hadron masses being not far from their vacuum
values. The line is almost independent of the parameter set and goes rather
close to the liquid-gas transition line.
The chiral crossover and the meson-baryon transition lines intersect
at $(T,\mu_B) \sim (150,450)$ MeV. 
The parity doublet model thus describes 3 domains: a chirally broken
phase with mesons thermodynamically dominating, another chirally broken phase 
where baryons are more dominant and the chirally restored phase,
which can be identified with quarkyonic matter~\cite{quarkyonic}.
It is worthy to note that this intersection point is fairly
close to the estimated triple point at which hadronic matter, quarkyonic matter
and quark-gluon plasma may coexist~\cite{triple}.

\section{Anomaly matching in matter}
\label{sec:amm}

How does deconfinement of colors enter to the chiral thermodynamics
at finite temperature and density? Although there exist a variety of 
studies using chiral Lagrangian approaches and holographic QCD models 
and a conjecture given in the large $N_c$ limit, no conclusive picture
on the actual QCD phase diagram is reached so far~\cite{triple}.

\begin{figure}[h]
\begin{center}
\includegraphics[width=12cm]{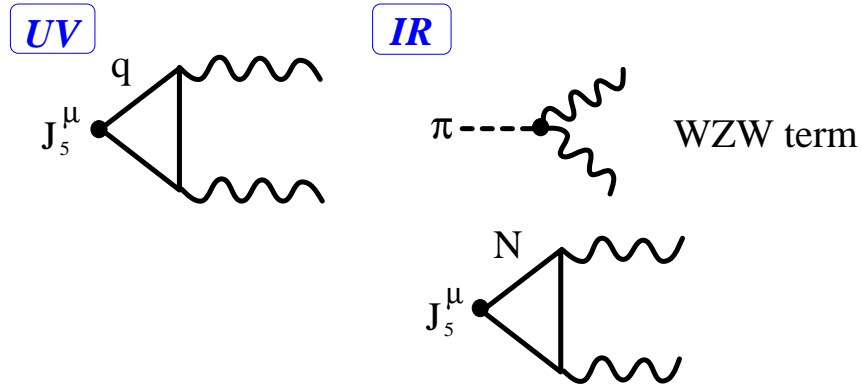}
\caption{
Saturations of the anomalies in terms of elementary quarks (left)
and of hadrons (right).
}
\label{anomaly}
\end{center}
\end{figure}

The anomaly matching is often used to constrain possible massless 
excitations in quantum field theories~\cite{thooft}.
External gauge fields, e.g. photons, interacting with quarks
lead to anomalies in the axial current (see Fig.~\ref{anomaly}). 
When the chiral symmetry is 
spontaneously broken in confined phase, the anomalies are saturated by 
the Nambu-Goldstone bosons. On the other hand, in chiral
restored phase, the anomalous contribution must be generated from the 
triangle diagram in which the baryons are circulating. In two flavors
the anomalies are matched with massless baryons, however, in three flavors,
the baryons forming an octet do not contribute to the pole in the 
axial current because of the cancellations~\cite{shifman:anomaly}.
Therefore, a system with restored chiral symmetry should be in deconfined
phase when the physics does not depend on the number of quark flavors
is imposed. For a system composed only from up and down quarks, the anomaly
matching does not exclude the chiral restored phase with confinement.
Nucleons with mirror assignment do not generate the anomalies 
since the axial couplings to the positive and negative parity states
have the same strength and their signs are relatively opposite.

\section{Summary and Discussions}
\label{sec:sum}

The parity doublet model within the mean field approximation describes 
the nuclear matter ground state at zero temperature and a chiral crossover 
at zero chemical potential at a reasonable temperature, which are the minimal 
requirements to describe the QCD thermodynamics. The first-order phase 
transitions appear only at low temperatures, below $T \sim 30$ MeV. 
Nevertheless, at higher temperature they still affect the order parameter 
which exhibits a substantial decrease near the liquid-gas {\it and} chiral 
transitions. If the chiral symmetry restoration is of first order, 
criticality around the end points of the two first-order phase transitions
will be the same due to the identical universality class~\cite{cp}.

We have also discussed the anomaly matching and possible model-selection.
In the presence of hot and dense matter the situation is more involved
due to the lack of Lorentz covariance and the existence of thermal 
masses~\cite{anomalymatching}. Besides, at high density gapless excitations 
on the Fermi surface might appear, which can be either bosons or fermions.
Since relevant low-energy excitations in matter are not perfectly known,
those degrees of freedom must be introduced in such a way that the model
certainly saturates the correct anomalies. The anomaly matching thus has
a role of the working hypothesis in modeling QCD matter.
It is indispensable to any rigorous argument for this taking account of 
the physics with medium effects, which could lead to a possibility 
of the chirally restored phase with confinement.

\begin{acknowledgments}
I am grateful for fruitful collaboration with I.~Mishustin.
Partial support by the Hessian
LOEWE initiative through the Helmholtz International
Center for FAIR (HIC for FAIR) is acknowledged.
\end{acknowledgments}

\newpage

\newpage

\begin{center}
FIGURE CAPTIONS
\end{center}
\begin{enumerate}
\item
The phase diagram in the parity doublet model~\cite{pdmour}.
The mass of the negative parity nucleon was taken to be $1.5$ GeV (left)
and $1.2$ GeV (right).
\item
Saturations of the anomalies in terms of elementary quarks (left)
and of hadrons (right).
\end{enumerate}


\newpage

\begin{thebibliography}{99}
\bibitem{qmproc}
C.~Sasaki,
  Nucl.\ Phys.\  A {\bf 830}, 649C (2009).

\bibitem{walecka}
  B.~D.~Serot and J.~D.~Walecka,
  Int.\ J.\ Mod.\ Phys.\  E {\bf 6}, 515 (1997).

\bibitem{other}
  J.~Boguta,
  Phys.\ Lett.\  B {\bf 120}, 34 (1983),
 I.~Mishustin, J.~Bondorf and M.~Rho,
  Nucl.\ Phys.\  A {\bf 555}, 215 (1993),
 G.~W.~Carter and P.~J.~Ellis,
  Nucl.\ Phys.\  A {\bf 628}, 325 (1998),
P.~Papazoglou, S.~Schramm, J.~Schaffner-Bielich, H.~Stoecker and W.~Greiner,
  Phys.\ Rev.\  C {\bf 57}, 2576 (1998),
P.~Papazoglou, D.~Zschiesche, S.~Schramm, J.~Schaffner-Bielich, H.~Stoecker and W.~Greiner,
  Phys.\ Rev.\  C {\bf 59}, 411 (1999).

\bibitem{nnjl}
  V.~Koch, T.~S.~Biro, J.~Kunz and U.~Mosel,
  Phys.\ Lett.\  B {\bf 185}, 1 (1987),
  M.~Buballa,
  Nucl.\ Phys.\  A {\bf 611}, 393 (1996),
  I.~N.~Mishustin, L.~M.~Satarov and W.~Greiner,
  Phys.\ Rept.\  {\bf 391}, 363 (2004).

\bibitem{bentz}
  W.~Bentz and A.~W.~Thomas,
  Nucl.\ Phys.\  A {\bf 696}, 138 (2001),
  W.~Bentz, T.~Horikawa, N.~Ishii and A.~W.~Thomas,
  Nucl.\ Phys.\  A {\bf 720}, 95 (2003).

\bibitem{dk}
  C.~E.~Detar and T.~Kunihiro,
  Phys.\ Rev.\  D {\bf 39}, 2805 (1989).

\bibitem{mirror}
  Y.~Nemoto, D.~Jido, M.~Oka and A.~Hosaka,
  Phys.\ Rev.\  D {\bf 57}, 4124 (1998),
  D.~Jido, Y.~Nemoto, M.~Oka and A.~Hosaka,
  Nucl.\ Phys.\  A {\bf 671}, 471 (2000),
  D.~Jido, T.~Hatsuda and T.~Kunihiro,
  Phys.\ Rev.\ Lett.\  {\bf 84}, 3252 (2000),
  D.~Jido, M.~Oka and A.~Hosaka,
  Prog.\ Theor.\ Phys.\  {\bf 106}, 873 (2001).

\bibitem{lsma1}
  S.~Gallas, F.~Giacosa and D.~H.~Rischke,
  Phys.\ Rev.\  D {\bf 82}, 014004 (2010).

\bibitem{hp}
  T.~Hatsuda and M.~Prakash,
  Phys.\ Lett.\  B {\bf 224}, 11 (1989).

\bibitem{pdm}
  D.~Zschiesche, L.~Tolos, J.~Schaffner-Bielich and R.~D.~Pisarski,
  Phys.\ Rev.\  C {\bf 75}, 055202 (2007).

\bibitem{astro}
  V.~Dexheimer, S.~Schramm and D.~Zschiesche,
  Phys.\ Rev.\  C {\bf 77}, 025803 (2008),
  V.~Dexheimer, G.~Pagliara, L.~Tolos, J.~Schaffner-Bielich and S.~Schramm,
  Eur.\ Phys.\ J.\  A {\bf 38}, 105 (2008).

\bibitem{pdmour}
  C.~Sasaki and I.~Mishustin,
  Phys.\ Rev.\  C {\bf 82}, 035204 (2010).

\bibitem{quarkyonic}
  L.~McLerran and R.~D.~Pisarski,
  Nucl.\ Phys.\  A {\bf 796}, 83 (2007),
  Y.~Hidaka, L.~D.~McLerran and R.~D.~Pisarski,
  Nucl.\ Phys.\  A {\bf 808}, 117 (2008),
  L.~McLerran, K.~Redlich and C.~Sasaki,
  Nucl.\ Phys.\  A {\bf 824}, 86 (2009).

\bibitem{triple}
  See, e.g. references in
  A.~Andronic {\it et al.},
  Nucl.\ Phys.\  A {\bf 837}, 65 (2010).

\bibitem{thooft}
G.~'t Hooft, in {\it Recent Developments in Gauge Theories},
ed. G.~'t Hooft {\it et al.} (Plenum Press, New York, 1980).

\bibitem{shifman:anomaly}
  M.~A.~Shifman,
  Phys.\ Rept.\  {\bf 209}, 341 (1991)
  [Sov.\ Phys.\ Usp.\  {\bf 32}, 289 (1989\ UFNAA,157,561-598.1989)].


\bibitem{cp}
  A.~M.~Halasz, A.~D.~Jackson, R.~E.~Shrock, M.~A.~Stephanov 
  and J.~J.~M.~Verbaarschot,
  Phys.\ Rev.\  D {\bf 58}, 096007 (1998).

\bibitem{anomalymatching}
H.~Itoyama and A.~H.~Mueller,
  Nucl.\ Phys.\  B {\bf 218}, 349 (1983),
  R.~D.~Pisarski,
  Phys.\ Rev.\ Lett.\  {\bf 76}, 3084 (1996),
  R.~D.~Pisarski, T.~L.~Trueman and M.~H.~G.~Tytgat,
  Phys.\ Rev.\  D {\bf 56}, 7077 (1997),
S.~D.~H.~Hsu, F.~Sannino and M.~Schwetz,
  Mod.\ Phys.\ Lett.\  A {\bf 16}, 1871 (2001),
  A.~Das and J.~Frenkel,
  arXiv:1012.1832 [hep-th].

\end{thebibliography}
\end{document}